\input{aipcheck}
\documentclass[final]{aipproc}
\layoutstyle{6x9}
\usepackage{amsmath}
\begin{document}
%
\title{Polaron in the $t$-$J$ Models with Three-Site Terms: \\
the $SU(2)$ and the Ising Cases}
\author{Krzysztof Wohlfeld}{address={Marian Smoluchowski Institute of Physics, Jagellonian University, \\Reymonta 4, PL-30059 Krak\'ow, Poland}}
\begin{abstract}
We compare the role of the three-site terms added to the $t$-$J$ models in the $SU(2)$ 
and the Ising cases in the extremely low doping regime, i.e. when a single hole added
to the strongly interacting half-filled system becomes a polaron. We show that in the realistic Ising case 
the three-site terms play a vital role in the polaron movement and should never be neglected unlike in the $SU(2)$ case. 
\end{abstract}
\maketitle
\section{INTRODUCTION}
The Hubbard model is regarded as $the$ model which can describe
many of the spectacular phenomena observed in transition metal
oxides (TMO) -- such as e.g. Mott insulating ground states, antiferromagnetism,
and perhaps even high temperature superconductivity \cite{Ima98}.
Usually, however, it is the $t$-$J$ model which is used to explain
these phenomena \cite{Ima98}. This model: 
($i$) is easier to solve than the Hubbard model \cite{HubN1},
($ii$) can be easily reduced to the Heisenberg model for the half-filled doping 
and hence naturally predicts the antiferromagnetic ground state of undoped TMO,
($iii$) is believed to give qualitatively similar predictions as the Hubbard model
for it merely follows from a perturbative expansion of the Hubbard model 
in the physical regime of strong interactions \cite{Cha78}.

A priori, one can have some doubts concerning the stated above $1\!:\!1$ correspondence
between these two models. Actually, in a rigorous perturbative expansion of the Hubbard
model one obtains the so-called three-site terms, in addition to the kinetic and
magnetic terms in the $t$-$J$ model \cite{Cha78}. These terms are often neglected in the 
small doping regime (close to the Mott insulating ground state) when they are much smaller 
than the two $t$-$J$ model terms and only yield a small longer range hopping. Indeed, 
it was shown in Ref. \cite{Bal95} that the  inclusion of the three-site terms in this 
standard $SU(2)$ symmetric case of the $t$-$J$ models does not change qualitatively 
the low-energy physics of the system.

However, a different situation can arise when the Ising limit of the $t$-$J$ model
is taken and the $SU(2)$ symmetry is broken, which is quite often done for computational purposes 
in e.g. more elaborate $t$-$J$ models \cite{Cos07}. Could the three-site
terms play an important role and entirely change the solutions in the Ising limit?
In this paper we answer this question by comparing the role of the three-site terms
added to the $t$-$J$ models in the $SU(2)$ and the Ising cases, cf. \cite{Dag08}. 
We perform calculations
in the extremely low doping regime namely for the case of a single hole added to 
a half-filled ordered ground state of the models, i.e. when a single hole couples
to the excitations of the ordered state in a polaronic way.

\section{$t$-$J$ models with three-site terms }
In order to see how the three-site terms arise in the perturbation theory and why they are often neglected 
we introduce (a generalized version of) the Hubbard model:
\begin{align} 
H=-t\sum_{\substack{i, j(i), \\ \sigma, \sigma' }} z_{ij}^{\sigma \sigma'} c^\dag_{i\sigma} c_{j\sigma'} + U\sum_i n_{i1}n_{i2},
\end{align}
where we defined $n_{i\sigma}=c^\dag_{i\sigma}c_{i\sigma}$, $\sigma=1, 2$ -- a quantum number with two values
(not necessarily a spin quantum number), and $z_{ij}^{\sigma \sigma'}$ is a hopping element between state $\sigma$ on site $i$
and state $\sigma'$ on site $j$. Hence, this model allows for e.g. $\sigma$-dependent values of the hopping elements. 
As already mentioned, usually in realistic systems the hopping $t$ is much smaller than the on-site interaction $U$ 
so that we can safely perform a perturbative expansion in the kinetic terms \cite{Cha78}. We obtain the effective Hamiltonian:
\begin{align}
H_{\mbox{\footnotesize eff}}&= 
-t\sum_{\substack{i, j(i), \\ \sigma, \sigma' }} 
\hat{t}_{ij}^{\sigma \sigma'} \tilde{c}^\dag_{i\sigma} \tilde{c}_{j\sigma'} 
-\frac{1}{4}J\sum_{\substack{i, \sigma_{1,...,4}, \\ j(i), j'(i)}}  
\big( z_{j'i}^{\sigma_4 \sigma_3} z_{ij}^{\sigma_2 \sigma_1} \tilde{c}^\dag_{j'\sigma_4} c_{i\sigma_3} n_{i\bar{\sigma}_3}n_{i\bar{\sigma}_2} c^\dag_{i\sigma_2} \tilde{c}_{j\sigma_1} \big) \\
&=\underbrace{-t\sum_{\substack{i, j(i), \\ \sigma, \sigma' }} \big( ... \big)-\frac{1}{4}J\sum_{\substack{i, \sigma_{1,...,4}, \\ j(i)=j'(i)}} \big( ... \big )}_{H_{t-J}}  \ \ \underbrace{-\frac{1}{4}J\sum_{\substack{i, \sigma_{1,...,4}, \\ j(i)\neq j'(i)}} \big( ... \big )}_{H_{3s}}, \label{eq:3}
\end{align}
where $\tilde{c}^\dag_{i\sigma}=c^\dag_{i\sigma}(1-n_{i\bar{\sigma}})$ and $J=4t^2/U$. 

The first two terms in Eq. (\ref{eq:3}) form a generalized version of the $t$-$J$ model ($H_{t-J}$) whereas the last one
is the three-site term ($H_{3s}$). The physical interpretation of these terms is the following: 
($i$) the first term describes the hopping of electrons with no double occupancies allowed and this term
contributes to the total energy of the system as $ \propto t \delta$,
where $\delta$ is the number of doped holes (by definition we have no holes in the half-filled case),
($ii$) the second term $\propto J {(1-\delta)}^2$ describes the electron which makes a virtual hopping $t$ from site $j$
to an occupied site $i$ which costs energy $U$ and then comes back to site $j$,
($iii$) the third term $\propto J \delta$ describes the situation when the electron does not return to the same site $j$ but goes
to the site $j'$ provided this site is unoccupied (hence in the half-filled case
this term vanishes).
Since in the low doping regime $\delta \ll 1$ and $J<t$, then naturally the last term
has the smallest contribution to the total energy of the system and this is the reason why 
(as explained in the introduction) it is often neglected. 

Let us also note the reasons for introducing the generalized version of the Hubbard and $t$-$J$ models.
Firstly, it enables us to treat the Ising and the $SU(2)$ cases on equal footing (see below).
Secondly, it is possible to obtain the Ising limit from this model in such a way that it coincides
with the model describing the correlated electrons in the systems with partial $t_{2g}$ orbital degeneracy
(with active $yz, zx$ orbitals in the ferromagnetic plane) as e.g. in the $d^1$ system of Sr$_{2}$VO$_4$ \cite{Noz91}. 
This means that this specific Ising limit would not only correspond to a standard Ising spin 
approximation $but$ would also constitute a realistic physical model \cite{Dag08}.

\section{Polaron in the $SU(2)$ case}
\begin{figure}[t]
\begin{minipage}{0.46\textwidth}
    \includegraphics[width=\textwidth]{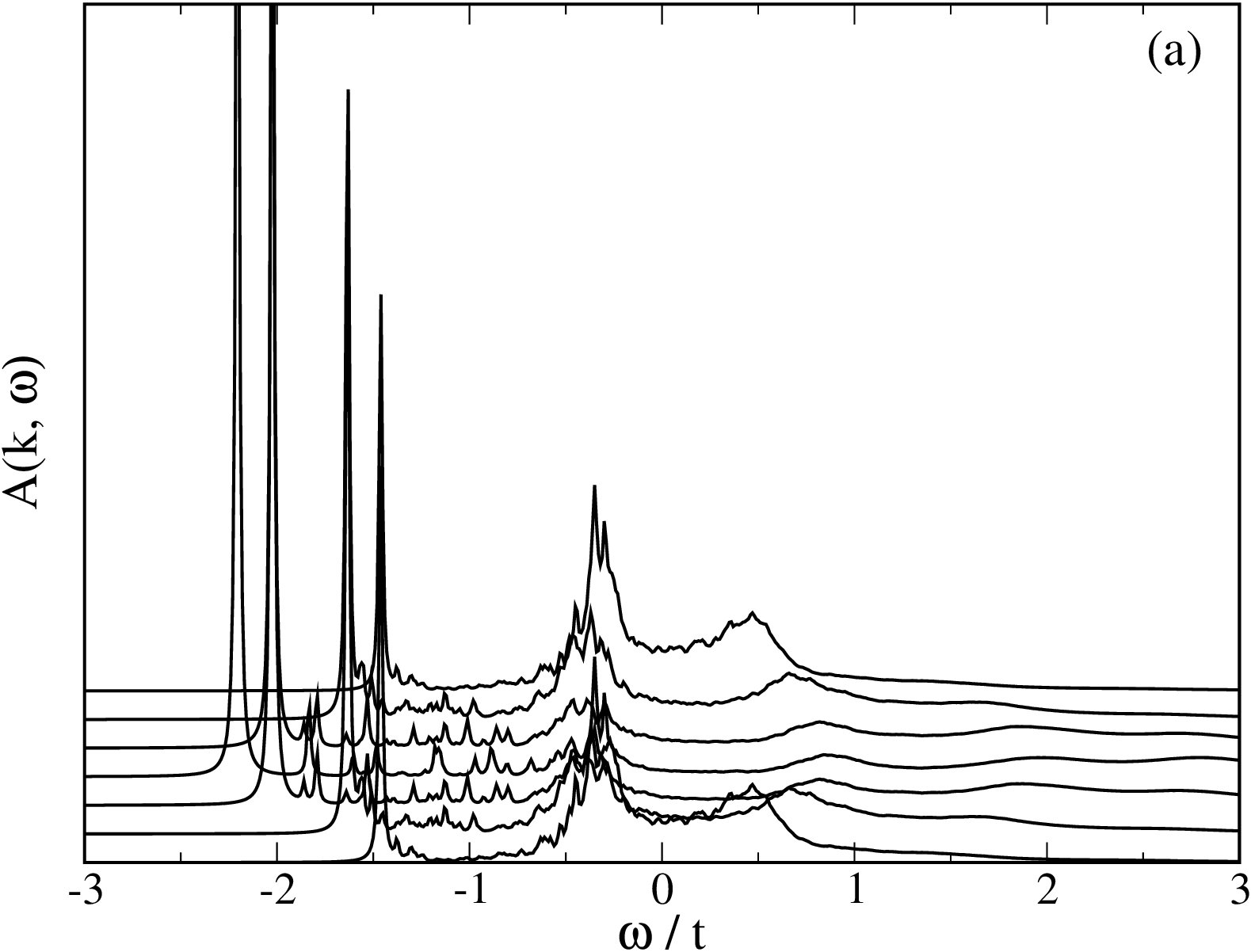}
\end{minipage}
\hfill \hfill \quad
\begin{minipage}{0.46\textwidth}
    \includegraphics[width=\textwidth]{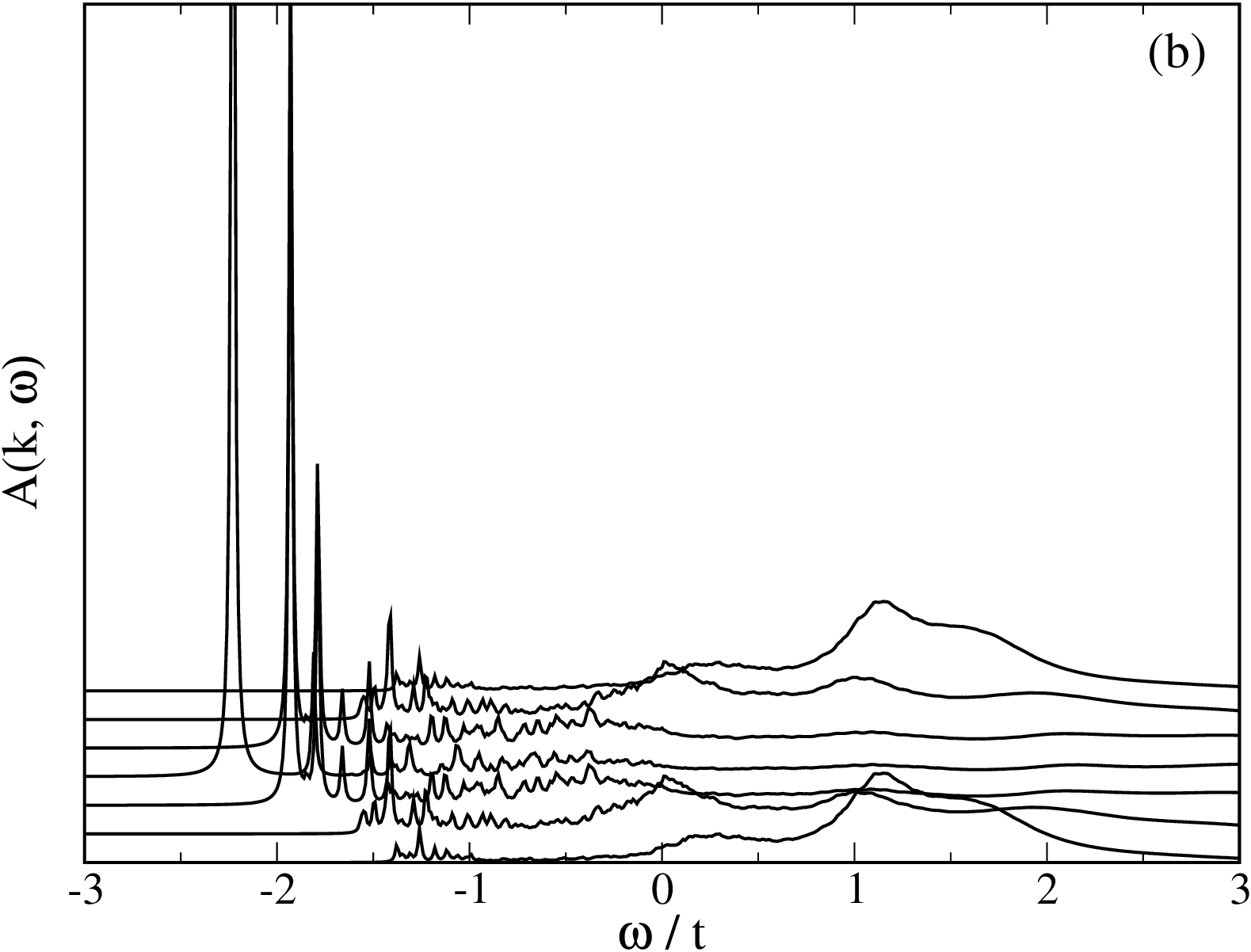}
\end{minipage}
\caption{\small{
Spectral density $A({\bf k}, \omega)$ of the model Eq. (\ref{eq:5}) with $J=0.4 t$ 
along the $\Gamma-\mbox{M}$ (from bottom to top) direction of the 2D Brillouin zone 
without (with) the first two terms of the three-site terms Eq. (\ref{eq:6}) 
shown on the left (right) panel, respectively.
} }
\label{fig:1}
\end{figure} 
In the $SU(2)$ symmetric case the quantum numbers $\sigma=\uparrow, \downarrow$
describe physical spins. Then
\begin{align}
z_{\langle ij \rangle}^{\sigma \sigma'}=\left( \begin{array}{cc}
1 & 0 \\
0 & 1 
\end{array} \right),
\label{eq:4}
\end{align}
where the hopping is restricted to the nearest neighbour sites $\langle i j\rangle$. Substituting Eq. (\ref{eq:4})
to Eq. (\ref{eq:3}) we obtain for the $t$-$J$ part
\begin{align}
&H_{t-J}=-t\sum_{\langle i j\rangle} (\tilde{c}^\dag_{i\uparrow} \tilde{c}_{j\uparrow}+\tilde{c}^\dag_{i\downarrow} \tilde{c}_{j\downarrow}+h.c.)+J\sum_{\langle i j\rangle}({\bf S}_i {\bf S}_{j}-\frac{1}{4}\tilde{n}_i\tilde{n}_{j}),
\label{eq:5}
\end{align}
where $S^+_i=\tilde{c}^\dag_{i\uparrow}\tilde{c}_{i\downarrow}$, $S^-_i=\tilde{c}^\dag_{i\downarrow}\tilde{c}_{i\uparrow}$, $S^z_i=\frac{1}{2}(\tilde{n}_{i\uparrow}-\tilde{n}_{i\downarrow})$. The three-site terms part reads
\begin{align}
H_{3s}=
-\frac{1}{4}J\sum_{\substack{\{j'ij\}}} (\tilde{c}^\dag_{j'\uparrow} \tilde{n}_{i \downarrow} \tilde{c}_{j \uparrow}
+\tilde{c}^\dag_{j'\downarrow} \tilde{n}_{i\uparrow} \tilde{c}_{j\downarrow}
-\tilde{c}^\dag_{j'\uparrow}  \tilde{c}^\dag_{i\downarrow} \tilde{c}_{i\uparrow} \tilde{c}_{j\downarrow} 
-\tilde{c}^\dag_{j' \downarrow }  \tilde{c}^\dag_{i \uparrow } \tilde{c}_{i \downarrow } \tilde{c}_{j \uparrow}  ),
\label{eq:6}
\end{align}
where we sum over all possible configurations of three adjacent sites $\{j'ij\}$ (with $i$ denoting the middle site).

In the half-filled case it is easily seen that the ground state of the model is a quantum antiferromagnet (AF).
The calculation of the ground state of the model with one additional hole in the AF is more complicated and one needs to use 
the self-consistent Born approximation (SCBA) \cite{Mar91}. 
The hole spectral density $A({\bf k}, \omega)$ obtained in this way for the two-dimensional (2D) lattice is 
shown in Fig. \ref{fig:1}(a) [Fig. \ref{fig:1}(b)] without [with] three-site terms  included, respectively. 
In both figures we see a finite though strongly renormalized dispersion of the lowest peak. 
This sugests a formation of a $mobile$ $polaron$ as a hole moves by dressing up with spin excitations
and acquires a large but finite effective mass. The inclusion of the three-site terms Eq. (\ref{eq:6}) 
merely strongly increases the spectral weight of the incoherent part
for particular ${\bf k}$ values in the Brillouin zone, cf. Fig. \ref{fig:1}(b). 
\section{Polaron in the Ising case}
\begin{figure}[t]
\begin{minipage}{0.46\textwidth}
    \includegraphics[width=\textwidth]{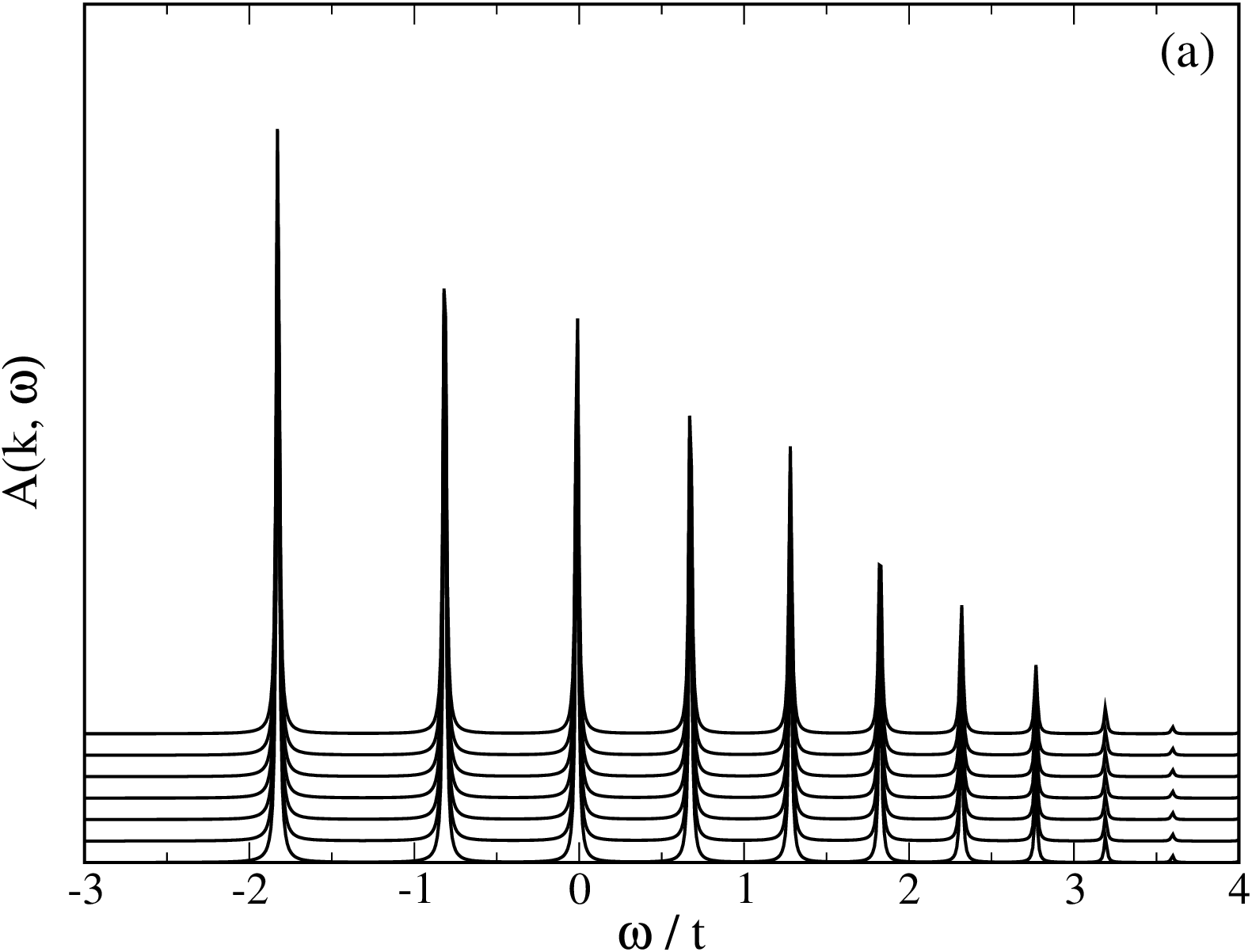}
\end{minipage}
\hfill \hfill \quad
\begin{minipage}{0.46\textwidth}
    \includegraphics[width=\textwidth]{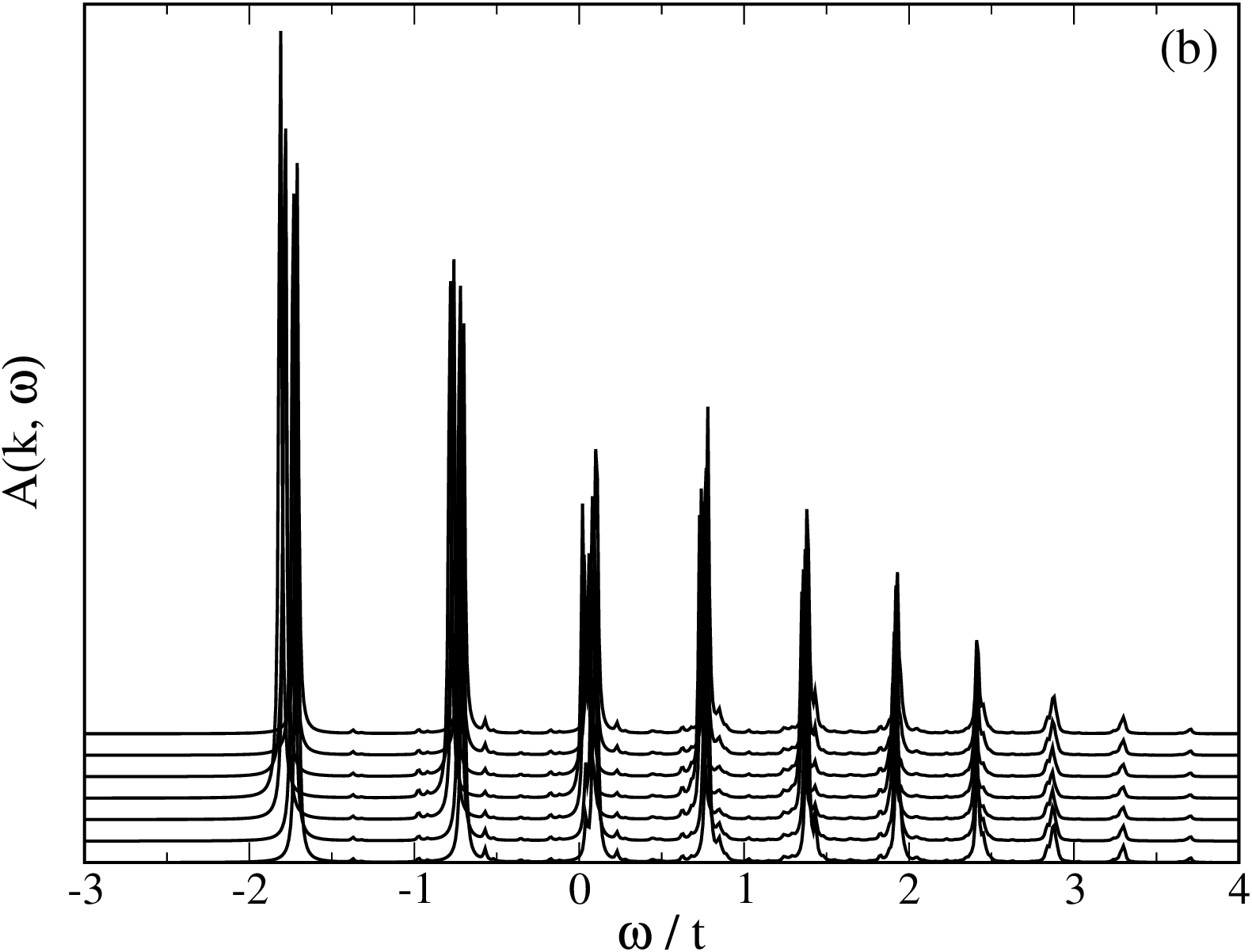}
\end{minipage}
\caption{\small{
Spectral density $A({\bf k}, \omega)$ of the model Eq. (\ref{eq:8}) with $J=0.4 t$ 
along the $\Gamma-\mbox{M}$ direction (from bottom to top) of the 2D Brillouin zone 
without (with) the first two terms of the three-site terms Eq. (\ref{eq:9}) shown on the left (right) panel, respectively.}}
\label{fig:2}
\end{figure} 
To obtain the Ising limit from the model Eq. (\ref{eq:3}) which at the same time serves as the realistic model
for correlated electrons with $t_{2g}$ degeneracy (see last paragraph of the second section) 
we assume $\sigma=yz, zx$ and
\begin{align}
z_{\langle i j \rangle || \hat{x}}^{\sigma \sigma'}=\left( \begin{array}{cc}
0 & 0 \\
0 & 1 
\end{array} \right)
\quad \mbox{and} \quad
z_{\langle i j \rangle || \hat{y}}^{\sigma \sigma'}=\left( \begin{array}{cc}
1 & 0 \\
0 & 0 
\end{array} \right),
\label{eq:7}
\end{align}
where the hopping is between the nearest neighbour sites along the $\hat{x}$ and $\hat{y}$ directions depending on the value of $\sigma$ \cite{Dag08}.
Substituting Eq. (\ref{eq:7}) to Eq. (\ref{eq:3}) we obtain for the $t$-$J$ part
\begin{align}
&H_{t-J}=-t\sum_{\langle i j \rangle || \hat{x}}  (\tilde{c}^\dag_{ib} \tilde{c}_{jb}+h.c.)-t\sum_{\langle i j \rangle || \hat{y}}  (\tilde{c}^\dag_{ia} \tilde{c}_{ja}+h.c.)+J\sum_{\langle i, j \rangle }(T^z_i T^z_{j}-\frac{1}{4}\tilde{n}_i\tilde{n}_{j}),
\label{eq:8}
\end{align}
where $T^z_i=\frac{1}{2}(\tilde{n}_{ia}-\tilde{n}_{ib})$. The three-site terms part reads
\begin{align}
H_{3s}=&-\frac{1}{4}J\sum_{\substack{\{j' i j\} || \hat{x}}} \tilde{c}^\dag_{j'b} \tilde{n}_{ia} \tilde{c}_{jb}-\frac{1}{4}J\sum_{\substack{\{j' i j\} || \hat{y}}} \tilde{c}^\dag_{j'a} \tilde{n}_{ib} \tilde{c}_{ja} \nonumber \\
&+\frac{1}{4}J\sum_{\substack{\{j' i j\} || [\hat{y}, \hat{x} ] } } \tilde{c}^\dag_{j'a}  \tilde{c}^\dag_{ib} \tilde{c}_{ia} \tilde{c}_{jb}
+\frac{1}{4}J\sum_{\substack{\{j' i j\} || [\hat{x}, \hat{y} ] }} \tilde{c}^\dag_{j'b}  \tilde{c}^\dag_{ia} \tilde{c}_{ib} \tilde{c}_{ja},
\label{eq:9}
\end{align}
where the three-site hop can be either straight (along $\hat{x}$ or $\hat{y}$ direction) or 
along the corner [e.g. first hop along the $\hat{x}$ direction and then along the $\hat{y}$ one, cf. the last term
of Eq. (\ref{eq:9})]. 

In the half-filled case the ground state of the model is a $classical$ state with alternating orbitals occupied
on each site (Ising AF in the spin language). The ground state of the model with one additional hole in the 
Ising AF can again be calculated using the SCBA. The hole spectral densities $A({\bf k}, \omega)$ for the 2D
case are shown in Fig. \ref{fig:2} with and without three-site terms included.
We see that in the case $without$ the three-site terms the polaron is $immobile$ \cite{Mar91} whereas $adding$ the three-site 
terms yields a small $dispersion$.

\section{Conclusions and final discussion}

In conclusion, we studied a problem of a single hole 
doped into the half-filled ground state of two different cases
of the $t$-$J$ model. On one hand, in the $SU(2)$ case
we noticed a quite big quantitative difference in the polaron
behaviour between the models with and without the three-site terms.
However, the polaron was always mobile and the inclusion
of the three-site terms did not change qualitatively the low-energy
physics of the system, cf. \cite{Bal95}. On the other hand,
in the Ising case the polaron became mobile only $after$ including
the three-site terms in the model. 

This means that for a $t$-$J$ model in the Ising limit to become more realistic
one should include the three-site terms even in the extremely low doping regime. 
In particular, one should also include these terms for the models describing
the $t_{2g}$ systems such as e.g. Sr$_{2}$VO$_4$ where the Ising limit arises
not as an approximation to the $SU(2)$ case but as a genuine 
$t$-$J$ model.

As a $postscriptum$ one can discuss the reason why the three-site terms
play any role in the Ising limit of the $t$-$J$ model despite the fact that 
they are very small in the low doping regime (as stated in the introduction
and explictly shown in the second section).
We suggest the following answer: if there is a strong competition
between two processes in the system (e.g. the magnetic interaction and the 
hole hopping in the $t$-$J$ model) then the neglected third term 
(e.g. the three-site hopping) can decide about the ground state of the system. 
This is because it is the difference between these two
larger terms in the Hamiltonian which matters and with which 
the magnitude of the neglected term should be compared.

\begin{theacknowledgments}
I would like to thank the organising committee of the course for their financial support.
I thank Andrzej M. Ole\'s, Maria Daghofer and Peter Horsch for the extremely 
fruitful discussion during the common work on this subject. I am also particularly 
grateful to Andrzej M. Ole\'s for his invaluable help and ideas. 
This work was supported in part by the Foundation for Polish Science (FNP) 
and by the Polish Ministry of Science under grant No. N202 068 32/1481.

\end{theacknowledgments}
\bibliographystyle{aipproc}   
\bibliography{sample}

\begin{thebibliography}{1}
\bibitem{HubN1}  The Hilbert space associated with the $t$-$J$ Hamiltonian is 
                 smaller than for the Hubbard one: for $N$ 
                 lattice sites the dimension of the Hilbert space
                 for the half-filled case is $~\big[\frac{N!}{(N/2)!(N/2)!}\big]^2$ 
                 for the Hubbard and $ 2^N $  for the $t$-$J$ model.
\end{thebibliography}

\end{document}